\documentclass[prl,preprintnumbers, twocolumn, superscriptaddress, amsmath, l66amssymb]{revtex4-1}

\usepackage{graphicx}
\usepackage{dcolumn}
\usepackage{bm}
\usepackage{verbatim}
\usepackage{CJK}
\usepackage{amsbsy} 
\usepackage{microtype}
\usepackage[normalem]{ulem}
\usepackage{tikz}
\usepackage{braket}
\usepackage{natbib}
\usepackage{appendix}
\usepackage{color,amsthm,amsmath,graphicx,bm,epsfig,float,siunitx,multirow,xr,enumerate,epstopdf,soul}
\usepackage[breaklinks=true,colorlinks=true,linkcolor=blue,urlcolor=blue,citecolor=blue]{hyperref}
\usepackage[T1]{fontenc}

\def\HIM{Helmholtz-Institut, GSI Helmholtzzentrum für Schwerionenforschung, Mainz 55128, Germany}
\def\JGU{Johannes Gutenberg-Universität Mainz, Mainz 55128, Germany}
\def\Berkeley{Department of Physics, University of California, Berkeley, CA 94720-7300, USA}
\def\LDQP{State Key Laboratory of Low-Dimensional Quantum Physics, Department of Physics, Tsinghua University, Beijing 100084, China}
\def\BUAAISOE{School of Instrumentation Science and Opto-electronics Engineering, Beihang University, Beijing, 100191, China}
\def\BUAAHZ{Hangzhou Innovation Institute, Beihang University, Hangzhou, 310051, China}
\def\BUAARIFS{Hangzhou Extremely Weak Magnetic Field Major Science and Technology Infrastructure Research Institute, Hangzhou, 310051, China}
\def\JLU{Key Laboratory of Geophysical Exploration Equipment, Ministry of Education of China, Jilin University, Changchun 130012, China}
\def\CQI{Frontier Science Center for Quantum Information, Beijing, China}
\def\JU{Institute of Theoretical Physics, Jagiellonian University, Łojasiewicza 11, 30-348 Kraków, Poland}

\usepackage{makeidx}
\begin{document}
\title{Constraints on Spin-Spin-Velocity-Dependent Interaction}
\author{Wei Ji}
\affiliation{\HIM}\affiliation{\LDQP}\affiliation{\JGU}
\author{Weipeng Li}
\affiliation{\LDQP}
\author{Pavel Fadeev}
\affiliation{\HIM}\affiliation{\JGU}
\author{Filip Ficek}
\affiliation{\JU}
\author{Jianan Qin}
\affiliation{\HIM}\affiliation{\JLU}
\author{Kai Wei}
\email{weikai@buaa.edu.cn}
\affiliation{\BUAAISOE}\affiliation{\BUAAHZ}\affiliation{\BUAARIFS}
\author{Yong-Chun Liu}
\email{ycliu@tsinghua.edu.cn}
\affiliation{\LDQP}\affiliation{\CQI}
\author{Dmitry Budker}
\affiliation{\HIM}\affiliation{\JGU}\affiliation{\Berkeley}
\date{\today}
\begin{abstract}
The existence of exotic spin-dependent forces may shine light on new physics beyond the Standard Model. We utilize two iron shielded SmCo$_5$ electron-spin sources and two optically pumped magnetometers to search for exotic long-range spin-spin-velocity-dependent force. The orientations of spin sources and magnetometers are optimized such that the exotic force is enhanced and common-mode noise is effectively subtracted. We set direct limit on proton-electron interaction in the force range from 1\,cm to 1\,km. Our experiment represents more than ten orders of magnitude improvement than previous works.
\end{abstract}
\maketitle


 The nature of dark matter is one of the most profound mysteries in modern physics. Many new light bosons introduced by theories beyond the Standard Model are proposed to be dark matter candidates, such as spin-0 bosons including axions and axion like particles (ALPs)\,\cite{axion-PhysRevLett.38.1440,preskill1983cosmology,graham2015experimental}, spin-1 bosons including dark photons\,\cite{darkPhoton.2010, Dark-Photon-AN2015331}, and Z$'$ bosons\,\citep{okada2020dark,croon2021supernova}. Furthermore, the new bosons may mediate new types of  long-range fundamental forces\,\citep{moody1984new,dobrescu2006spin,Fadeev2019,fayet1986fifth}.

If we consider the spin, relative position and velocity of two fermions, the exotic interaction between them can be classified to 16 terms  \cite{dobrescu2006spin,Fadeev2019}, and then generally classified into static terms and velocity-dependent terms. A conventional velocity-dependent force in classical physics is the Lorentz force of a moving charged particle.

Many experimental methods have been used to  search for exotic forces, including experiments with torsional resonators\,\cite{heckel2008preferred,hammond2007new,Loong-Nature2003,terrano2015short}, nuclear magnetic resonance\,\cite{petukhov2010polarized,yan2015searching,chu2013laboratory,arvanitaki2014resonantly}, magnetometers based on hot atoms and nitrogen-vacancy center in diamond\,\cite{Romalis-PRL2009,Ji2018,jiao2021experimental,hunter2014using,su2021search,almasi2020new,kim2018experimental}, and other high-sensitivity technologies\,\cite{tullney2013constraints,serebrov2010search,ficek2017constraints,ren2021search,stadnik2018improved}.  Most of these efforts focus on static interactions, while the velocity-dependent interactions have also been gaining attention in recent years\,\cite{jiao2021experimental,su2021search,hunter2014using,wei2022new,kim2018experimental}.

In this experiment, we focus on one term of Spin-Spin-Velocity-Dependent Interaction (SSVDI) proposed by\,\cite{dobrescu2006spin}:
\begin{align}
\label{eq.v8}
V=&\frac{f \hbar}{4\pi c} \left[(\hat{\boldsymbol\sigma}_1 \cdot \boldsymbol{v})(\hat{\boldsymbol\sigma}_2 \cdot\boldsymbol{v})\right]\ \frac{e^{-r/\lambda}}{r},
\end{align}
where $f$ is a dimensionless coupling coefficient, 
 $\hat{\boldsymbol\sigma}_1$, $\hat{\boldsymbol\sigma}_2$  are the respective Pauli spin-matrix vectors of the two fermions,
$r$ and $\mathbf{v}$ are the relative position and velocity between two fermions. (This potential is called $V_8$ in Ref.\,\cite{dobrescu2006spin}.) To search for this force, a spin polarized test object is required as the spin source, and an ultra-sensitive magnetometer is required as the sensor. 

In this experiment, the spin sources are two iron shielded SmCo$_5$ magnets (ISSCs) that have high net electron spin and small magnetic leakage \cite{ji2017searching}. The sensor is a pair of optically pumped magnetometers (OPM) that operate in the spin-exchange-relaxation-free (SERF) mode\,\cite{allred2002high,shah2007subpicotesla}, which use spin polarized Rb as the sensing atoms. By designing the setup, the experiment is sensitive to the exotic force, while common-mode noise is reduced. Our experiment sets new limits on exotic SSVDI for electron-proton coupling.

\begin{figure}
\begin{center}
\includegraphics[width=9cm]{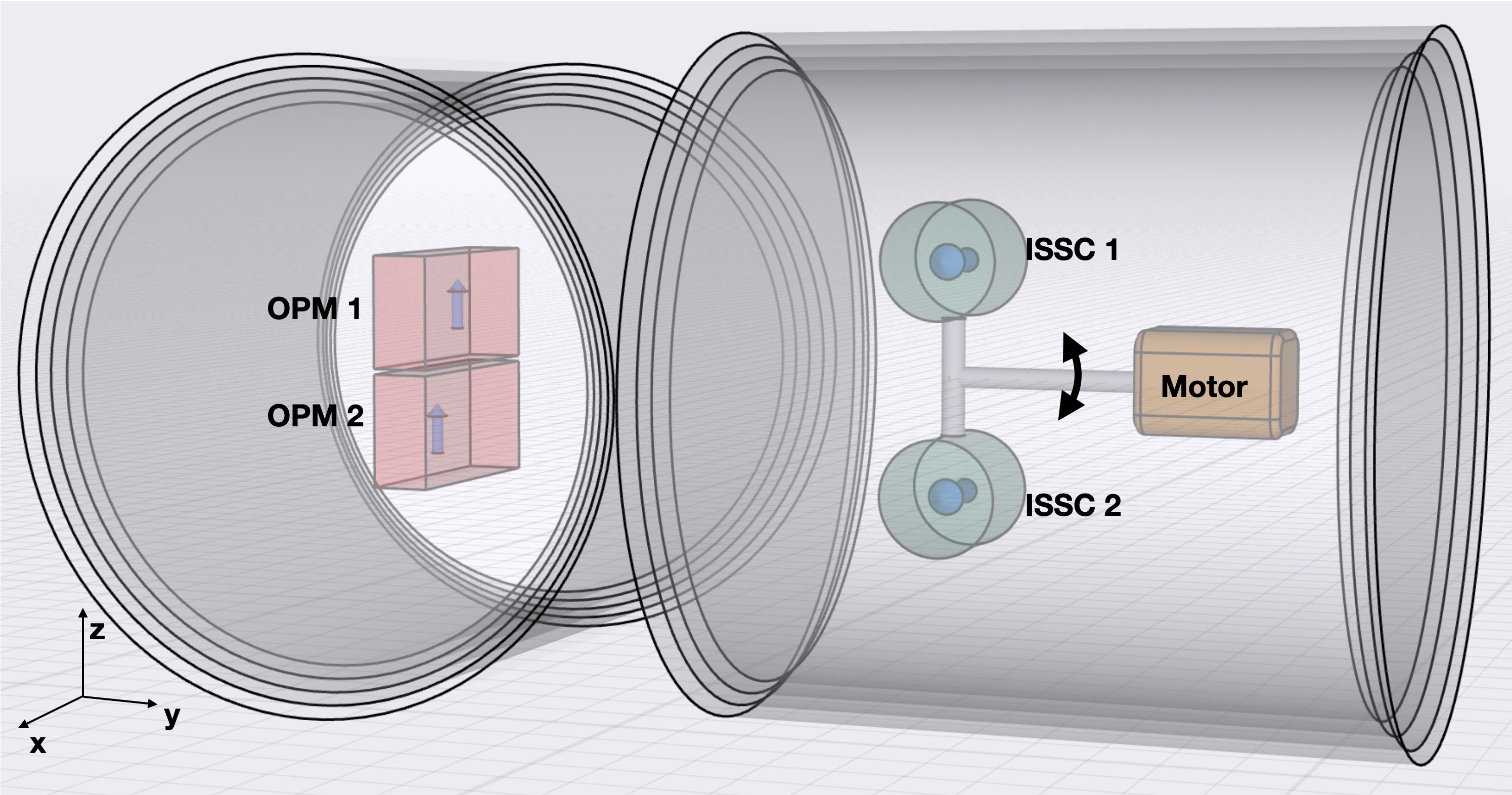}
\caption{
 The experimental setup (not to scale). Two QuSpin OPMs noted as OPM1 and OPM2 are enclosed in a five-layer magnetic shield. Their sensitive axis orientations are antiparallel along the $\hat{x}$-axis.
 Two spin sources noted as ISSC$_{1,2}$ are put in the other, four-layer shield. The spin source is driven with a motor to rotate clockwise or counterclockwise. The blue arrows show the direction of net spin in OPMs and ISSCs.
}
\label{Fig.Exp.Setup} 
\end{center}
\end{figure}

Figure\,\ref{Fig.Exp.Setup} shows a schematic of the experimental setup. Each of the two spin sources ISSC$_{1,2}$ contains a 40.00 mm diameter cylindrical SmCo$_5$ magnet enclosed in three layers of pure iron. 
The magnetization of the SmCo$_5$ magnets is about 1\,T. The magnetic field of the magnet is shielded by the iron layers, and the magnetic leakage outside the iron layers is smaller than 10\,$\mu$T. However, the net spin of the ISSCs is not canceled, which is mostly due to the fact that the orbital magnetic moment and spin magnetic moments of the 4$f$ rare earth metal (Sm) and 3$d$ metals (Co and Fe) are differently oriented, and thus the total magnetic moments are canceled but the net orbital magnetic moment and spin magnetic moment are not \cite{ji2017searching,heckel2008preferred}. 
 The net electron spin for each ISSC is 1.75 $(21)\times 10^{24}$ \cite{ji2017searching}. The ISSCs are connected with titanium-alloy supports and are driven with a motor to rotate clockwise (CW) and counterclockwise (CCW). The motor frequency is controlled with a direct current (DC) power supply. 
 
The OPMs are QuSpin Vector Zero-Field Magnetometers (QZFM Gen-2)\,\cite{Quspin} that work in the SERF regime. They are placed in the center of a five-layer $\mu$-metal magnetic shield. The arrows along the $\hat{z}$-axis demonstrate the direction of the circularly polarized laser beam that passes through the $^{87}$Rb vapor cell. A narrow linewidth Rb Hannle resonance in near-zero field is used to detect the magnetic field\,\cite{dupont1969detection}.  
 Because the orientation of two OPMs along $\hat{x}$ are anti-parallel, their responses to the magnetic field have opposite signs. If there is a magnetic field $B_0$ applied, the responses of the OPMs are $S_1=B_0+N_{C}+N_{1}$ and $S_2=-B_0+N_C +N_{2}$, respectively, where $N_{C}$ is the common-mode noise and $N_{1}$ and $N_{2}$ are other noises. Subtracting the readings of the two sensors can diminish the common noise and yields a signal of $S_{\mathrm{sub}}=(S_1-S_2)/2=B_0+(N_1-N_2)/2$. 
 
\begin{figure}
\begin{center}
\includegraphics[width=9.5cm]{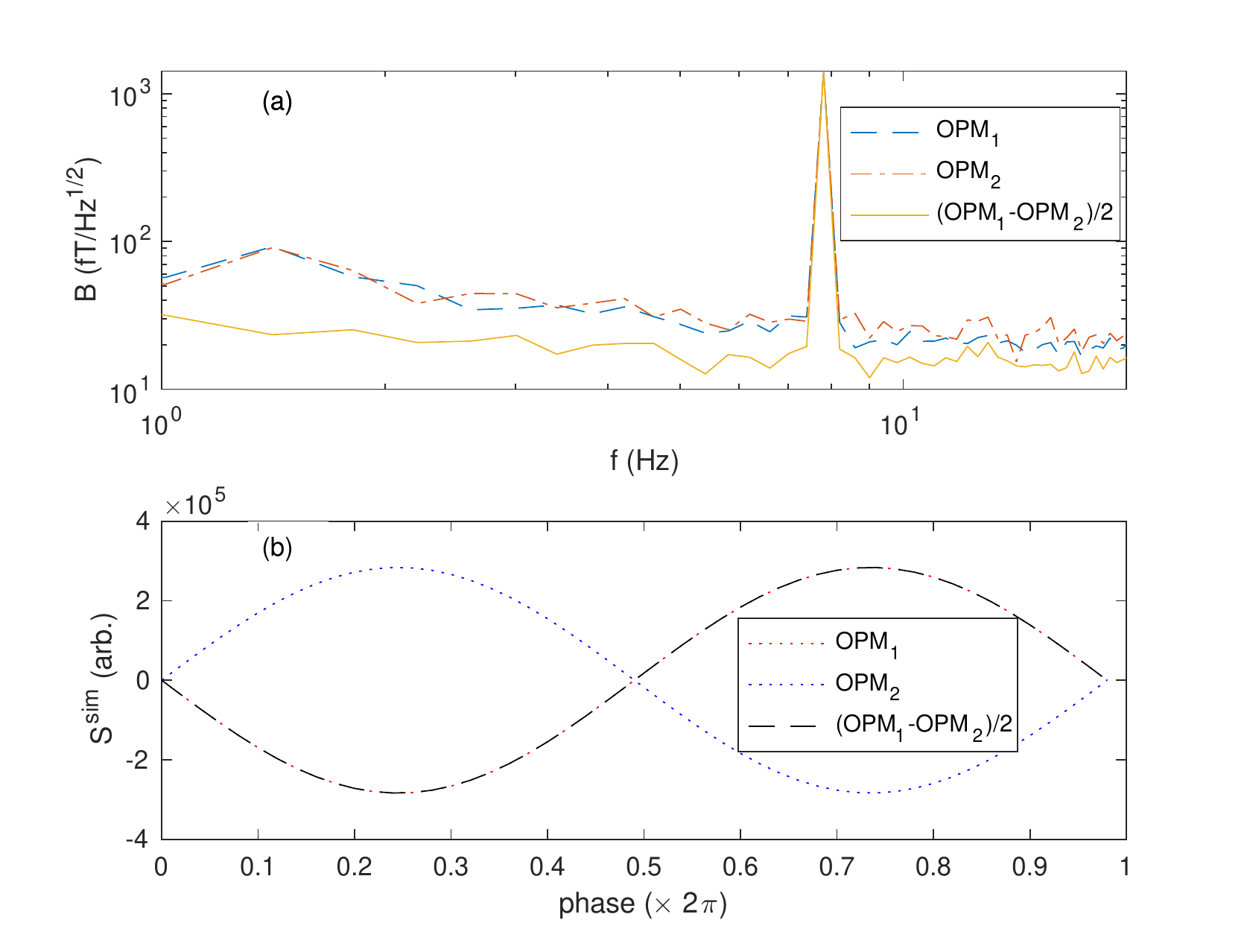}
\caption{
{Up: A typical spectrum of two OPMs and the subtraction result. A uniform AC magnetic field of 8\,Hz is applied along the $\hat{x}$-axis. The dashed-blue line and the red-dot-dashed line are the spectrum of the OPMs on the left side and right side, respectively. The yellow-solid line is their difference.
Down: The OPMs' response to the pseudomagnetic field along the $\hat{x}$-axis. The blue-dot and red-dot line are the pseudomagnetic field sensed by OPM$_1$ and OPM$_2$ respectively, the black-dash line is the subtraction result. The subtraction result agrees well with the result from OPM$_1$.} 
}
\label{fig.power.spectrum.simulation}  
\end{center}
\end{figure}

 To test the validity of the subtraction procedure, an 8\,Hz and 1.5 pT uniform magnetic field is applied along $\hat{x}$ with a set of Helmholtz coils. The spectrum of the OPM signals and the subtraction result are shown in Fig.\,\ref{fig.power.spectrum.simulation}\,(a). By taking the difference, the uniform magnetic field is unaffected, the common-mode (for example, electrical or gradient) noise is reduced by as high as a factor of five, and the 8\,Hz target signal is successfully extracted. The noise level around 8\,Hz is about 13\,fT/$\sqrt{\textrm{Hz}}$. 


The SSVDIs will manifest as a pseudomagnetic fields that could be sensed by the Rb atoms like the Zeeman effect. The potential can be expressed as $V^n\zeta^n+V^p\zeta^p+V^e\zeta^e= -\boldsymbol{\mu} \cdot \textbf{B}$, where $\boldsymbol{\mu}$ is the magnetic moment of the Rb atom, $\textbf{B}$ is the pseudomagnetic field from the exotic interaction, $\zeta^{n,p,e}$ are the neutron, proton and electron's fraction of spin polarization in $^{87}$Rb atoms, which could be obtained by the the Russel-Saunders LS-coupling and the Schmidt model of nuclear physics.

In this experiment, we search for the coupling between the proton, neutron and electron spins in the Rb atoms and the electron spins in ISSCs. The pseudomagnetic field sensed by the OPM can be obtained by integrating the exotic interaction from the electron spins over the ISSCs:
\begin{align}
\begin{split}
\mathbf{B}^{p,e,n}=&\frac{ f \zeta^{p,e,n} \hbar}{4\pi \mu c}
\iiint \rho(\mathbf{r})
(\hat{\boldsymbol\sigma}_2\cdot \mathbf{v})
\frac{\mathbf{v} }{r}e^{-r/\lambda}
\rm{d}\mathbf{r},
\end{split}
\label{eq.beff}
\end{align}
where $\textbf{B}^{p,n,e}$ are the fractions of $\textbf{B}$ that couple to proton, neutron and electron respectively, $\mathbf{v}(\mathbf{r})=\mathbf{\omega}\times\mathbf{r}$ and 
$\rho(\mathbf{r})$ are the velocity and spin density at location $\mathbf{r}$, $\mathbf{\omega}$ is the angular velocity of the ISSCs. The proton and electron fractions of polarization in $^{87}$Rb are $\zeta^p=0.29$ and $\zeta^e=0.13$ respectively, and neutron polarization $\zeta^n$ is assumed to be zero under the basic nuclear shell model. The calculation of the fraction of spin polarization is explained in the supplemental document.

The experimental parameters and a benchmark coupling coefficients $f ^{0}=1$ are put in the simulation to obtain $\textbf{B}^p$. The benchmark parameter $f ^{0}$ is set to $1$ for convenience; a different $f ^{0}$ does not affect the final result.
The orientation of the OPMs and ISSC sources are optimized by simulating different configurations, such that the OPMs can sense the maximum pseudomagnetic field. The best configuration is shown in Figure\,\ref{Fig.Exp.Setup} and Table\,\ref{Tab.Parameters}. The distance between the spin sources and the OPMs is much larger than the distance between two OPMs, such that two OPMs experience almost the same pseudomagnetic field. Thus the signal subtraction procedure works well for this pseudomagnetic field. The simulated responses of the two OPMs and their subtraction result are shown in Fig.\,\ref{fig.power.spectrum.simulation}\,(b).

\begin{table}[!h]
\begin{ruledtabular}
\caption {{Experimental parameters and the error budget of $f^{\rm{ep}}$.}
The origin of coordinates is at the midpoint between the centers of the two OPMs. The contributions  to the error budget are evaluated for $\lambda=20$\,m. The final systematic error is derived from the uncertainties of the parameters listed.
}
\label{Tab.Parameters}
\begin{tabular}{c c c c} 
Parameter & Value & $\Delta f^{\rm{ep}}  (\times 10^{-{22}})$
  \\
ISSC net spin ($\times10^{24}$) &$1.75(21) $ & 0.084\\
Position of ISSCs x(m)& $0.000(2)$  & $0.001$\\
Position of ISSCs y(m)& $-0.477(2)$ & $0.001$\\ 
Position of ISSCs z(m)& $0.000(2) $  & $0.001$\\
Distance between ISSCs(m) & $0.251(1)$   & $0.044$\\
Distance between OPM cells(m) & $0.017(1)$   & $0.004$\\
\hline
Rotation frequency CW (Hz) &$4.11(1)$  & \\
Rotation frequency CCW (Hz) &$4.09(1)$  & \\
phase uncertainty ($\deg$) &$\pm 2.8 $& $\pm 1.190$\\
\hline
Final $f^{\rm{exp}}  (\times 10^{-22}$) & $-0.7$
 & $\pm 10.1\ (stat.)$\\
$(\lambda=20\,m)$ &  & $\pm 1.2 (syst.)
$ 
\end{tabular} 
\end{ruledtabular}
\end{table}

The ISSC spin sources are driven with a DC motor. The positions of the spin sources are monitored with a photoelectronic encoder placed on the rotation axle. The signals of the encoder and the OPMs are taken simultaneously and recorded with a data-acquisition (DAQ) device. The motor is tuned to rotate CCW and CW alternatively for every two hours. The DC motor works in a good stability with frequency of $4.09 (1)$\,Hz and $4.11 (1)$\,Hz for CW and CCW rotations. 

The two OPMs signals are subtracted and then transformed to frequency domain by fast Fourier transformations (FFT). The 50 Hz power line interference and its 100 Hz and 200 Hz harmonics are removed in the frequency domain. The data were then transformed back to the time domain with inverse FFT.   

The signals are then cut to one-period-long segments based on the encoder signal of the spin source rotation. The DC components in each period are removed. The data are noted as $\mathbf{S}^{\rm{exp}}_{i}(t_{j})$, where $i$ represents the $i$-th period, and $t_j$ is the time of the $j$-th point in this period.

The coupling coefficient $f^{\rm{ep}}$ can be obtained by a similarity comparison method between the experimental data and simulation results: 
\begin{equation}
f^{\rm{ep}}_{i}=k_i\, \sqrt{\frac{\sum_j \left[\mathbf{S}^{\rm{exp}}_i(t_j)\right]^2}
{\sum_j \left[\mathbf{S}^{\rm{sim}}(t_j)\right]^2}} \, ,
\label{eq.fi}
\end{equation}
where $k_i$ is the similarity score to weigh the similarity between  $\mathbf{S}^{\rm{exp}}_i$ and $\mathbf{S}^{\rm{sim}}(t)$\,\cite{krasichkov2015shape}, which is defined as 
\begin{equation}\label{eq.ki}
k_i
\equiv\frac{\sum_j \mathbf{S}^{\rm{sim}}(t_j)\cdot {\mathbf{S}}_i^{\rm{exp}}(t_j)}
{\sqrt{\sum_j \left[\mathbf{S}^{\rm{sim}}(t_j)\right]^2}
\sqrt{\sum_j \left[{\mathbf{S}_i}^{\rm{exp}}(t_j)\right]^2}
}.
\end{equation}

The expectation values and standard error for the CW and CCW rotation are $\langle f^{\rm{ep}}\rangle^{+}$,  $\langle f^{\rm{ep}}\rangle^{-}$, and $\sigma_f^+$, $\sigma_f^-$ respectivly, 
the final coupling coefficient can be obtained by
\begin{equation}\label{eq.average}
\langle f^{\rm{ep}}\rangle =
\frac{\langle f^{\rm{ep}}\rangle^{+}/{\sigma^+}^2+ \langle f^{\rm{ep}}\rangle^{-}/{\sigma^-}^2}
{1/{\sigma^+}^2+1/{\sigma^-}^2} 
.
\end{equation}

\begin{figure}
\begin{center}
\includegraphics[width=9.cm]{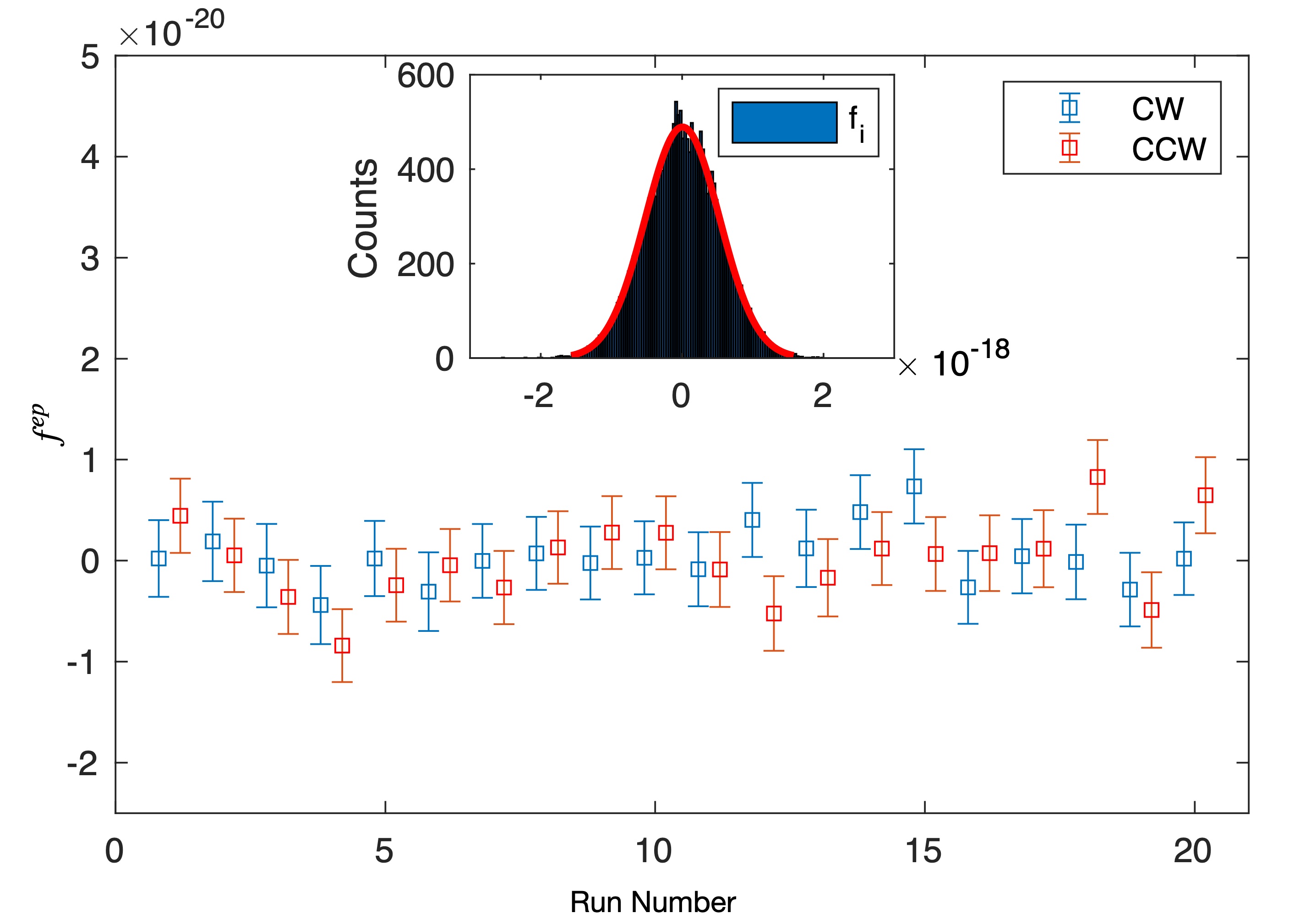}
\caption{Statistical results of the $f ^{\rm{ep}}$. Each data point represents an average of about one 2.7-hour long data set. The distribution of $f ^{\rm{ep}}$ for one data set is shown in the insert. The result is well fitted with a Gaussian distribution (red line) with $\bar{\chi}^2=1.18$. 
}
\label{Fig.Dist} 
\end{center}
\end{figure}

Some systematic bias could be removed by averaging over CW an CCW. The distributions of the $f^{\rm{ep}+}$ and $f^{\rm{ep}-}$ are shown in Fig.\,\ref{Fig.Dist}.

\begin{figure}
\begin{center}
 \includegraphics[width=8.5cm]{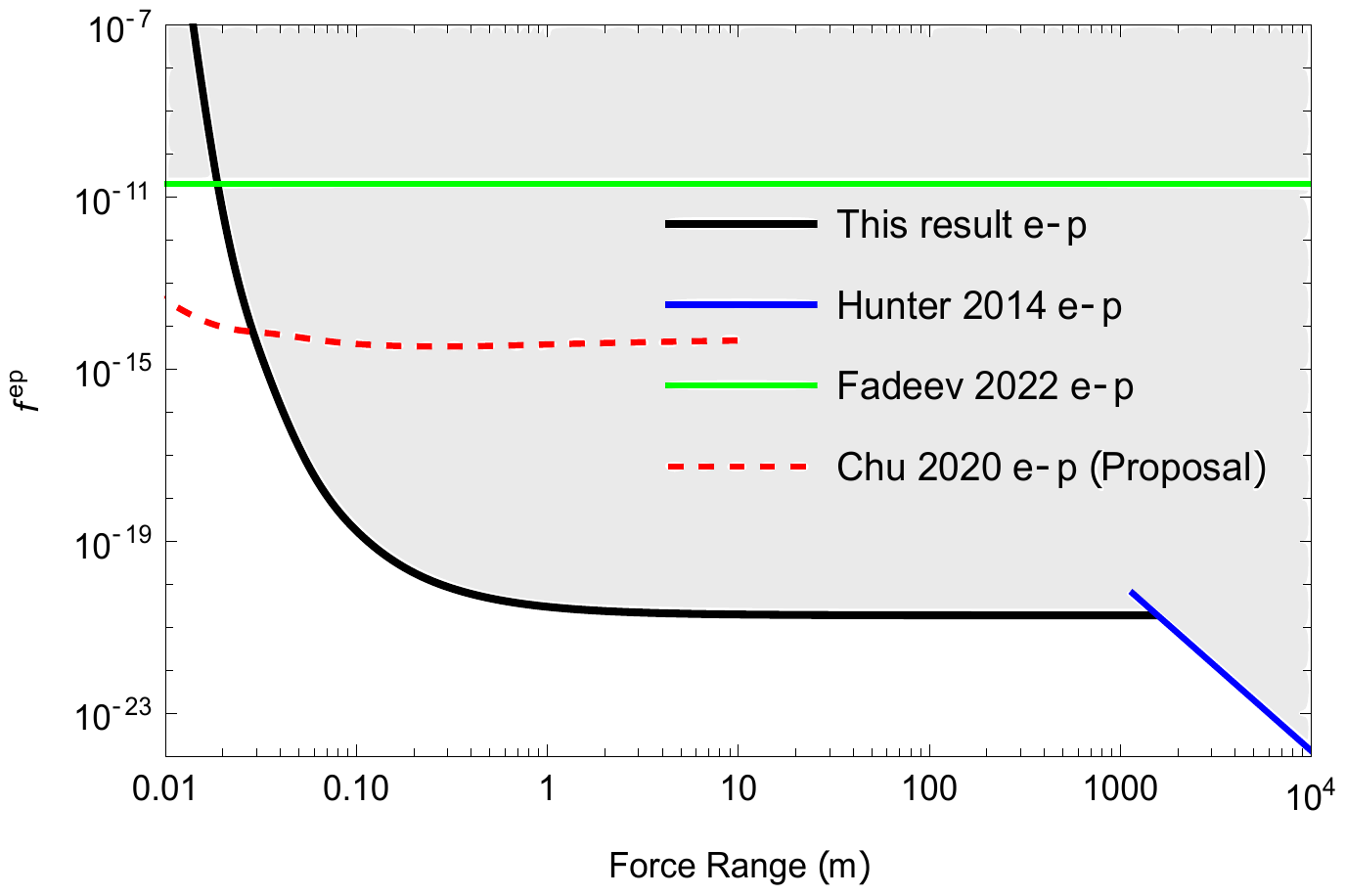}
\caption{
Limits on the SSVDI coupling coefficients between electron and proton. The black solid line is our constraints. The ``Hunter2014 e-p'' is from Ref.\,\cite{hunter2014using} that uses geo-electrons and atomic magnetometer; the ``Fadeev 2022 e-p'' is from Refs.\,\cite{Fadeev2022Pseudovector} that compare the experimental and theoretical results of hydrogen spectroscopy; the ``Chu 2020 e-p''\,\cite{chu2020search} propose  to use $^3$He as sensor and dysprosium iron garnet as spin source, the line is based on their sensitivity at $3 \times 10^{-17}$ T and is rescaled using the fraction of spin polarisation $\zeta_p^{^3 \text{He}}=-0.027$ \cite{vasilakis2009limits}.}
 \label{fig.results}  
\end{center}
\end{figure}


The parameters of the experiment and their corresponding uncertainties on $\Delta f ^{\rm{exp}}$ for range $\lambda=20$\,m are shown in Table\,\ref{Tab.Parameters}. The $f ^{\rm{ep}}$ is determined to be\,$ f ^{\rm{ep}}=0.7\pm10.1_{stat.}\pm 1.2_{syst.} (\times 10^{-22})$.
No evidence of the SSVDI is observed. New constraints on the $f $ between electron-proton is set to be $|f^{\rm{ep}}|\le 2.0\times 10^{-21}$ by the 95\% confidence level. The values for other $\lambda$s are obtained with the same procedure, and the final limits are shown in Fig. \,\ref{fig.results}.

If the mediator of the SSVDI is a spin-1 boson such as Z$'$, which is a dark matter candidate and may resolve other discr\rm{ep}ancies such as that in the anomalous magnetic moment of the muon\,\cite{arushi2021solving,okada2020dark}, the coupling coefficient can be rewritten as $f^{\rm{ep}}=-g_A^eg_A^p/2$\,\cite{dobrescu2006spin,Fadeev2019}.
For $\lambda=20$\,m, $|g_A^eg_A^p|\le4.0\times 10^{-21}$, where to set the limit on one of these coupling-constant products, we assume that the other one is zero.
Note that the velocity-independent term provides significantly tighter limit on $g_Ag_A$ coefficients\,\cite{Fadeev2019,dobrescu2006spin}, however, the SSVDI provides a unique way to explore the velocity-dependent interactions.

A comparison between our results (black and dashed-red lines) and the literature is shown in Fig.\,\ref{fig.results}. 
With the same hydrogen-spectrum analysis used in Ref.\,\cite{Fadeev2022Pseudovector}  we obtained a bound on the SSVDF of $f^{ep}<2.0\times 10^{-11}$ for the range larger than 1\,cm (the green line ``Fadeev 2022 e-p'' in Fig.\,\ref{fig.results}). Results on the couplings between other fermions, such as the coupling between electron-electron\,\cite{ficek2017constraints,hunter2014using,Ji2018} neutron-proton\,\cite{kimball2010constraints} and electron-antiproton\,\cite{ficek2018constraints} are not plotted on Fig.\,\ref{fig.results}. 

The major advance of our experiment is that the ISSC spin sources have much larger numbers of spins compared to those in precision-spectroscopy experiments yielding data for the analyses in \cite{ficek2017constraints,ficek2018constraints} and spin-exchange approaches\,\cite{kimball2010constraints}, which are most sensitive to forces with ranges on the atomic to microscopic scale. The other advantage is that the OPMs typically have energy resolution on the order of $10^{-18}$\,eV\,\cite{kimball2015nuclear}, significantly better than for the spectroscopy used in Refs.\cite{ficek2017constraints,ficek2018constraints}. 
On the other hand, spectroscopy experiments have an advantage over macroscopic once in the short range, because of the exponential decay of the exotic force.
Our search covers the range of parameters inaccessible for the geoelectron experiment\,\cite{hunter2014using}. Using the same method and data, we also set limits for the electron-proton coupling on the V$_{6+7}$, V$_{15}$ and V$_{16}$ terms of SSVDF \cite{dobrescu2006spin,Fadeev2019}. The results are shown in the supplemental document.

A major concern in this experiment was magnetic leakage from the ISSCs.
With the iron shielding, at a distance of 10\,cm away from the ISSC's surface, 
its residual magnetic field was measured to be less than 10\,$\mu$T. 
The shielding factors for the magnetic shielding of the ISSCs and OPMs were measured to be both greater than $10^{6}$. Considering all the decay and shielding factors, we conservatively expect the magnetic leakage from the ISSCs to the position of the OPMs to be smaller than 0.1 aT, which was insignificant with regards to the error budget.

The stability of the OPM is monitored throughout the experiment. The DC drift of the OPM is less than 2\,pT within two hours. A servo motor has a better frequency precision, however, commercial servo motor's control systems have electromagnetic coupling with the magnetometer \cite{Ji2018}. A DC motor is chosen to diminish this coupling. The experiment can further be improved if a larger size ISSC could be used.  The vapor cell can also be replaced with a magnetometer that uses a levitated ferromagnetic sphere and has orders of magnitude better potential magnetic sensitivity\,\cite{vinante2021surpassing}.


In summary, we utilized a pair of OPMs that can reduce the common noise and have ultrahigh sensitivity to search for exotic spin-dependent physics. Together with the high electron spin density iron-shielded SmCo$_5$ spin source, the new experiment sets new limits on SSVDI, with more than 10 orders of magnitude improvement for the electron-proton coupling.

\begin{acknowledgments}
We thank Dr. Derek F. Jackson Kimball and Dr. Changbo Fu for valuable discussions. This work is supported by Key-Area Research and Development Program of Guangdong Province (Grant No. 2019B030330001), the National Natural Science Foundation of China (NSFC) (Grant Nos. 12275145, 92050110, 91736106, 11674390, and 91836302), and the National Key R\&D Program of China (Grants No. 2018YFA0306504), the DFG Project ID 390831469: EXC 2118 (PRISMA+ Cluster of Excellence), by the German Federal
Ministry of Education and Research (BMBF) within the Quantumtechnologien program (Grant No. 13N15064).
and by the QuantERA project LEMAQUME (DFG Project Number 500314265).
\end{acknowledgments}

\bibliographystyle{naturemag}
\bibliography{5thForce}
\end{document}